\documentstyle[sprocl]{article}

\font\eightrm=cmr8

\def\Journal#1#2#3#4{{#1} {\bf #2}, #3 (#4)}

\def\NCA{\em Nuovo Cimento}
\def\NIM{\em Nucl. Instrum. Methods}
\def\NIMA{{\em Nucl. Instrum. Methods} A}
\def\NPB{{\em Nucl. Phys.} B}
\def\PLB{{\em Phys. Lett.}  B}
\def\PRL{\em Phys. Rev. Lett.}
\def\PRD{{\em Phys. Rev.} D}
\def\ZPC{{\em Z. Phys.} C}

\def\st{\scriptstyle}
\def\sst{\scriptscriptstyle}
\def\mco{\multicolumn}
\def\epp{\epsilon^{\prime}}
\def\vep{\varepsilon}
\def\ra{\rightarrow}
\def\ppg{\pi^+\pi^-\gamma}
\def\vp{{\bf p}}
\def\ko{K^0}
\def\kb{\bar{K^0}}
\def\al{\alpha}
\def\ab{\bar{\alpha}}
\def\be{\begin{equation}}
\def\ee{\end{equation}}
\def\bea{\begin{eqnarray}}
\def\eea{\end{eqnarray}}
\def\CPbar{\hbox{{\rm CP}\hskip-1.80em{/}}}

\begin{document}

\title{NON-PERTURBATIVE GLUODYNAMICS OF HIGH ENERGY HEAVY-ION COLLISIONS}

\author{A. KRASNITZ}

\address{UCEH, Universidade do Algarve \\
Campus de Gambelas, P-8000 Faro, Portugal \\
E-mail: krasnitz@ualg.pt}

\author{R. VENUGOPALAN}

\address{Physics Department, Brookhaven National Laboratory,\\
Upton, NY 11973, USA \\
E-mail: raju@bnl.gov}

\maketitle\abstracts{The dynamics of low-x partons in the transverse
plane of a high-energy nuclear collision is classical, and therefore
admits a fully non--perturbative numerical treatment. We report
results of a recent study estimating the initial energy density in the
central region of a collision. Preliminary estimates of the number of
gluons per unit rapidity, and the initial transverse momentum
distribution of gluons, are also provided.}

In heavy-ion experiments, planned at RHIC later this year, gold ions
are expected to collide at $\sqrt{s}=200$ GeV per nucleon.  A few
years later $\sqrt{s}=5.5$ TeV per nucleon will be attained in heavy
ion collisions at LHC.  In the central region of these collisions, a
combination of very high center-of-mass energy with a very large
number of participating valence quarks will likely give rise to a
novel regime of QCD, one characterized by a very high parton
density. This regime does not easily lend itself to a description
based on conventional approaches. Collisions involving large
transverse momenta can be adequately described in terms of pairwise
scattering of individual partons comprising the colliding
systems. Final-state interactions of secondary partons formed therein
can be safely neglected~\cite{KajMuell}.  However, as the parton
density grows, final-state interactions of secondary partons must be
taken into account. This requirement is only partially satisfied by
multiple scattering or by classical cascade descriptions, which ignore
the coherence of the secondary field configuration~\cite{mscl}.

The coherence of the secondary partons is incorporated naturally
into the classical effective field theory approach of McLerran and Venugopalan 
(MV)~\cite{RajLar}.
If the parton density in the colliding nuclei is high at small $x$, classical
methods are applicable. It has been shown recently that a RG-improved
generalization of this effective action reproduces several key results
in small-$x$ QCD: the leading $\alpha_s\log(1/x)$ BFKL equation, the double log
GLR equation and its extensions, and the small-$x$ DGLAP equation for quark
distributions~\cite{JKMW}. 

Briefly, the model is based on the following assumptions. Partons in a
nucleus are separated into high-$x$ and the low-$x$ components. The
former corresponds to valence quarks and hard sea partons. These
high-$x$ partons are considered recoilless sources of color charge.
For a large Lorentz-contracted nucleus, this results in a static
Gaussian distribution of their color charge density $\rho$ in the
transverse plane:
$$P\left([\rho]\right)\propto\, {\rm exp}\left[-{1\over{2g^4\mu^2}}
\int{\rm d}^2r_t\rho^2(r_t)\right].$$ The variance $\mu^2$ of the
color charge distribution is the only dimensional parameter of the
model, apart from the linear size $L$ of the nucleus. For central
impact parameters, $\mu$ is given in terms of single-nucleon structure
functions~\cite{GyulassyMclerran}:
$$\mu^2={A^{1/3}\over{\pi r_0^2}}\int_{x_0}^1{\rm
d}x\left({1\over{2N_c}} q(x,Q^2)+{N_c\over{N_c^2-1}}g(x,Q^2)\right),$$
with the separation scale $x_0\equiv Q/\sqrt{s}$, $r_0=1.12$ fm, and
$N_c$ the number of colors.  It is assumed, in addition, that the
nucleus is infinitely thin in the longitudinal direction. Under this
simplifying assumption the resulting gauge fields are boost-invariant.

The small $x$ fields are then described by the classical Yang-Mills equations
\begin{equation}
D_\mu F_{\mu\nu}=J_\nu\label{eqmo}\end{equation}
with the random sources on the two light 
cones:
$J_\nu=\sum_{1,2}\delta_{\nu,\pm}\delta(x_\mp)\rho_{1,2}(r_t).$
The two signs correspond to two possible directions of motion along the
beam axis $z$. As shown by Kovner, McLerran and Weigert (KMW), low $x$ fields
in the central region of the collision obey sourceless Yang-Mills equations
(this region is in the forward light cone of both nuclei) with the initial 
conditions in the $A_\tau=0$ gauge given by
\begin{equation}
A^i=A^i_1+A^i_2;\ \ \ \ A^\pm=\pm{{ig}\over 2}x^\pm[A^i_1,A^i_2].
\label{incond}\end{equation}
Here the pure gauge fields $A^i_{1,2}$ are solutions of (\ref{eqmo}) for each
of the two nuclei in the absence of the other nucleus. 

\begin{figure}[hbt]
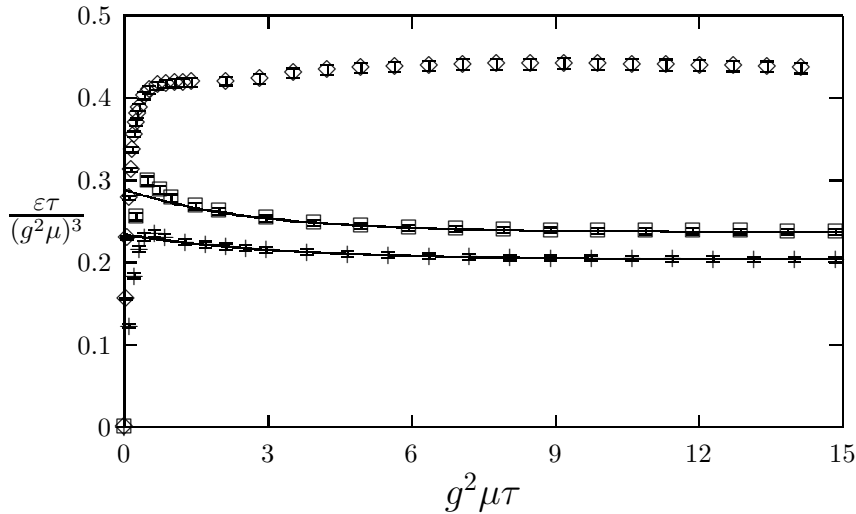

\setlength{\unitlength}{0.240900pt}
\ifx\plotpoint\undefined\newsavebox{\plotpoint}\fi
\sbox{\plotpoint}{\rule[-0.200pt]{0.400pt}{0.400pt}}%

\caption{Transverse-plane energy density per unit rapidity versus proper time
for the values 5.66 (diamonds), 35.36 (plusses), and 297 (squares) of $g^2\mu L$.
Both the energy density and the proper time are expressed in units of $g^2\mu$.
The solid lines are fits of the data to the form 
$\alpha+\beta\,{\rm exp}(-\gamma\tau)$.}
\label{ehist}\end{figure}
Equation (\ref{eqmo}) with the initial condition (\ref{incond}) can now be 
solved, in order to obtain the resulting gluon field configuration at late
proper times. Since the initial condition depends on the random color source,
averages over realizations of the source must be performed. This aspect of the
solution resembles the classical thermal theory, wherein an average is performed
over initial conditions drawn from the canonical ensemble. In fact, the analogy
between the MV effective theory and the classical thermal theory goes further.
This analogy can be made explicit by considering the perturbative solution of
(\ref{eqmo}) obtained by KMW. They showed that in perturbation theory
the gluon number distribution by transverse momentum (per unit rapidity)
suffers from an  infrared divergence and argued that the distribution must
have the form
\begin{equation}
n_{k_\perp}\propto{1\over\alpha_s}\left({{\alpha_s\mu}\over k_\perp}\right)^4\ln
\left({k_\perp\over{\alpha_s\mu}}\right) \label{dpt}\end{equation}
for $k_\perp\gg\alpha_s\mu$. We can now draw a parallel between 
$\mu$ and the temperature $T$ of the thermal system. In 
particular, 
the log term suggests that the perturbative description breaks down 
for $k_\perp\sim\alpha_s\mu$. Likewise, the perturbative thermal theory
loses validity at the non-perturbative scale $k\sim g^2T$.

It is therefore clear that a fully non--perturbative study of MV model
is necessary. The model is discretized on a lattice in
the transverse plane and the lattice field equations solved
numerically. Boost invariance and periodic boundary conditions in 
the transverse plane are assumed. Technical details of the lattice
formulation can be found in Ref.~\cite{AlexRaj}. 
The quantity $g^2\mu$ and the linear size $L$ of the nucleus
are the only physically interesting dimensional parameters of the MV
model~\cite{RajGavai}.  Any dimensional quantity $q$ can then be
written as $(g^2\mu)^df_q(g^2\mu L)$, where $d$ is the dimension of
$q$. All the non-trivial physical information is contained in the
dimensionless function $f_q(g^2\mu L)$. On a lattice, $q$ will
generally depend also on the lattice spacing $a$; we will seek to
remove this dependence by taking the continuum limit $a\rightarrow
0$. Finally, we estimate the values of the dimensional parameter
$g^2\mu L$ which correspond to key collider experiments. Assuming
Au-Au collisions, we take $L=11.6$ fm (for a square nucleus!) and estimate 
the standard
deviation $\mu$ to be 0.5 GeV for RHIC and 1 GeV for 
LHC~\cite{GyulassyMclerran}.  Also, we
have approximately $g=2$ for energies of interest. The rough estimate is
then $g^2\mu L\approx 120$ for RHIC and $g^2\mu L\approx 240$ for LHC.
Clearly, there is some variation in $g^2\mu L$ due to the various 
uncertainities in this estimate. The expression we will derive is a 
non--perturbative formula, from which one can deduce the number or 
energy of produced gluons for a particular choice of $g^2\mu L$.

\begin{figure}[hbt]
\setlength{\unitlength}{0.240900pt}
\ifx\plotpoint\undefined\newsavebox{\plotpoint}\fi
\sbox{\plotpoint}{\rule[-0.200pt]{0.400pt}{0.400pt}}%
\begin{picture}(1350,810)(0,0)
\font\gnuplot=cmr10 at 10pt
\gnuplot
\sbox{\plotpoint}{\rule[-0.200pt]{0.400pt}{0.400pt}}%
\put(161.0,215.0){\rule[-0.200pt]{4.818pt}{0.400pt}}
\put(141,215){\makebox(0,0)[r]{0.2}}
\put(1269.0,215.0){\rule[-0.200pt]{4.818pt}{0.400pt}}
\put(161.0,400.0){\rule[-0.200pt]{4.818pt}{0.400pt}}
\put(141,400){\makebox(0,0)[r]{0.3}}
\put(1269.0,400.0){\rule[-0.200pt]{4.818pt}{0.400pt}}
\put(161.0,585.0){\rule[-0.200pt]{4.818pt}{0.400pt}}
\put(141,585){\makebox(0,0)[r]{0.4}}
\put(1269.0,585.0){\rule[-0.200pt]{4.818pt}{0.400pt}}
\put(161.0,770.0){\rule[-0.200pt]{4.818pt}{0.400pt}}
\put(141,770){\makebox(0,0)[r]{0.5}}
\put(1269.0,770.0){\rule[-0.200pt]{4.818pt}{0.400pt}}
\put(161.0,123.0){\rule[-0.200pt]{0.400pt}{4.818pt}}
\put(161,82){\makebox(0,0){0}}
\put(161.0,750.0){\rule[-0.200pt]{0.400pt}{4.818pt}}
\put(337.0,123.0){\rule[-0.200pt]{0.400pt}{4.818pt}}
\put(337,82){\makebox(0,0){50}}
\put(337.0,750.0){\rule[-0.200pt]{0.400pt}{4.818pt}}
\put(514.0,123.0){\rule[-0.200pt]{0.400pt}{4.818pt}}
\put(514,82){\makebox(0,0){100}}
\put(514.0,750.0){\rule[-0.200pt]{0.400pt}{4.818pt}}
\put(690.0,123.0){\rule[-0.200pt]{0.400pt}{4.818pt}}
\put(690,82){\makebox(0,0){150}}
\put(690.0,750.0){\rule[-0.200pt]{0.400pt}{4.818pt}}
\put(866.0,123.0){\rule[-0.200pt]{0.400pt}{4.818pt}}
\put(866,82){\makebox(0,0){200}}
\put(866.0,750.0){\rule[-0.200pt]{0.400pt}{4.818pt}}
\put(1042.0,123.0){\rule[-0.200pt]{0.400pt}{4.818pt}}
\put(1042,82){\makebox(0,0){250}}
\put(1042.0,750.0){\rule[-0.200pt]{0.400pt}{4.818pt}}
\put(1219.0,123.0){\rule[-0.200pt]{0.400pt}{4.818pt}}
\put(1219,82){\makebox(0,0){300}}
\put(1219.0,750.0){\rule[-0.200pt]{0.400pt}{4.818pt}}
\put(161.0,123.0){\rule[-0.200pt]{271.735pt}{0.400pt}}
\put(1289.0,123.0){\rule[-0.200pt]{0.400pt}{155.862pt}}
\put(161.0,770.0){\rule[-0.200pt]{271.735pt}{0.400pt}}
\put(40,446){\makebox(0,0){\Large ${{\varepsilon\tau}\over{(g^2\mu)^3}}$}}
\put(725,21){\makebox(0,0){\Large $g^2\mu L$}}
\put(161.0,123.0){\rule[-0.200pt]{0.400pt}{155.862pt}}
\put(181,652){\raisebox{-.8pt}{\makebox(0,0){$\Diamond$}}}
\put(192,635){\raisebox{-.8pt}{\makebox(0,0){$\Diamond$}}}
\put(223,444){\raisebox{-.8pt}{\makebox(0,0){$\Diamond$}}}
\put(286,230){\raisebox{-.8pt}{\makebox(0,0){$\Diamond$}}}
\put(410,216){\raisebox{-.8pt}{\makebox(0,0){$\Diamond$}}}
\put(535,236){\raisebox{-.8pt}{\makebox(0,0){$\Diamond$}}}
\put(684,276){\raisebox{-.8pt}{\makebox(0,0){$\Diamond$}}}
\put(909,278){\raisebox{-.8pt}{\makebox(0,0){$\Diamond$}}}
\put(1208,321){\raisebox{-.8pt}{\makebox(0,0){$\Diamond$}}}
\end{picture}
\caption{Transverse-plane energy density (in units of $g^2\mu$) per unit 
rapidity versus $g^2\mu L$. The error bars are smaller than the plotting 
symbols.}
\label{esummary}\end{figure}
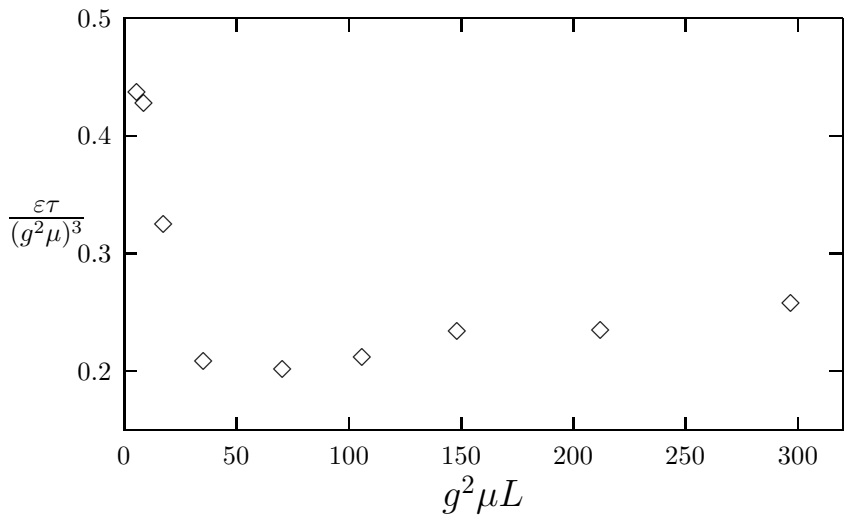
Results of a numerical investigation (for SU(2) only) are as follows. 
We first compute the energy per unit transverse area  per unit rapidity, 
deposited in the central region by the colliding nuclei. As Figure \ref{ehist}
illustrates, this quantity tends to a constant at late proper times. It is
this asymptotic value of the energy density that we wish to determine. 
If we express the energy density in units of $g^2\mu$ and extrapolate our
numerical findings to the continuum limit, we find that the energy density
depends on the dimensionless parameter $g^2\mu L$ as described in Figure
\ref{esummary}. Note the very slow variation of this dimensionless function in 
the entire range of $g^2\mu L$ values, which includes both our RHIC and LHC
estimates. Using this plot, and assuming, in accordance with 
Ref.~\cite{Muell2},
the $(N_c^2-1)/N_c$ dependence of the energy on the number of colors $N_c$,
we arrive at the values of 2700 GeV and of 25000 GeV for the transverse energy 
per unit rapidity at RHIC and at LHC, respectively~\cite{AlexRaj1}.

\begin{figure}[hbt]
\setlength{\unitlength}{0.240900pt}
\ifx\plotpoint\undefined\newsavebox{\plotpoint}\fi
\sbox{\plotpoint}{\rule[-0.200pt]{0.400pt}{0.400pt}}%
\begin{picture}(1350,810)(0,0)
\font\gnuplot=cmr10 at 10pt
\gnuplot
\sbox{\plotpoint}{\rule[-0.200pt]{0.400pt}{0.400pt}}%
\put(181.0,231.0){\rule[-0.200pt]{4.818pt}{0.400pt}}
\put(161,231){\makebox(0,0)[r]{0.12}}
\put(1269.0,231.0){\rule[-0.200pt]{4.818pt}{0.400pt}}
\put(181.0,662.0){\rule[-0.200pt]{4.818pt}{0.400pt}}
\put(161,662){\makebox(0,0)[r]{0.16}}
\put(1269.0,662.0){\rule[-0.200pt]{4.818pt}{0.400pt}}
\put(181.0,123.0){\rule[-0.200pt]{0.400pt}{4.818pt}}
\put(181,82){\makebox(0,0){0}}
\put(181.0,750.0){\rule[-0.200pt]{0.400pt}{4.818pt}}
\put(354.0,123.0){\rule[-0.200pt]{0.400pt}{4.818pt}}
\put(354,82){\makebox(0,0){50}}
\put(354.0,750.0){\rule[-0.200pt]{0.400pt}{4.818pt}}
\put(527.0,123.0){\rule[-0.200pt]{0.400pt}{4.818pt}}
\put(527,82){\makebox(0,0){100}}
\put(527.0,750.0){\rule[-0.200pt]{0.400pt}{4.818pt}}
\put(700.0,123.0){\rule[-0.200pt]{0.400pt}{4.818pt}}
\put(700,82){\makebox(0,0){150}}
\put(700.0,750.0){\rule[-0.200pt]{0.400pt}{4.818pt}}
\put(874.0,123.0){\rule[-0.200pt]{0.400pt}{4.818pt}}
\put(874,82){\makebox(0,0){200}}
\put(874.0,750.0){\rule[-0.200pt]{0.400pt}{4.818pt}}
\put(1047.0,123.0){\rule[-0.200pt]{0.400pt}{4.818pt}}
\put(1047,82){\makebox(0,0){250}}
\put(1047.0,750.0){\rule[-0.200pt]{0.400pt}{4.818pt}}
\put(1220.0,123.0){\rule[-0.200pt]{0.400pt}{4.818pt}}
\put(1220,82){\makebox(0,0){300}}
\put(1220.0,750.0){\rule[-0.200pt]{0.400pt}{4.818pt}}
\put(181.0,123.0){\rule[-0.200pt]{266.917pt}{0.400pt}}
\put(1289.0,123.0){\rule[-0.200pt]{0.400pt}{155.862pt}}
\put(181.0,770.0){\rule[-0.200pt]{266.917pt}{0.400pt}}
\put(40,446){\makebox(0,0){\Large ${N\over{(g^2\mu L)^2}}$}}
\put(735,21){\makebox(0,0){\Large $g^2\mu L$}}
\put(181.0,123.0){\rule[-0.200pt]{0.400pt}{155.862pt}}
\put(303,332){\raisebox{-.8pt}{\makebox(0,0){$\Diamond$}}}
\put(426,286){\raisebox{-.8pt}{\makebox(0,0){$\Diamond$}}}
\put(695,440){\raisebox{-.8pt}{\makebox(0,0){$\Diamond$}}}
\put(916,500){\raisebox{-.8pt}{\makebox(0,0){$\Diamond$}}}
\put(1209,567){\raisebox{-.8pt}{\makebox(0,0){$\Diamond$}}}
\put(303,166){\makebox(0,0){$+$}}
\put(426,221){\makebox(0,0){$+$}}
\put(695,570){\makebox(0,0){$+$}}
\put(916,516){\makebox(0,0){$+$}}
\put(1209,738){\makebox(0,0){$+$}}
\end{picture}
\caption{Transverse-plane gluon number density per unit rapidity versus 
$g^2\mu L$ from Coulomb gauge fixing (diamonds) and from relaxation (plusses).}
\label{numtot}\end{figure}
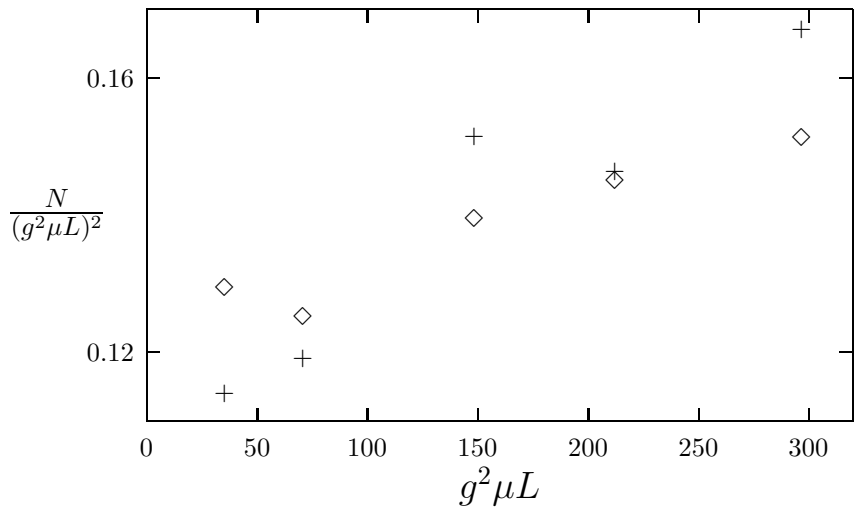
Next, we report our preliminary estimates of low-$x$ gluon
multiplicities in the central region. Determination of the total
number of produced gluons is of considerable interest: this quantity 
may be directly related to the 
number of produced hadrons~\cite{McLerran}. Further, the
momentum distribution of gluons in the transverse plane can be used as
initial data for a Boltzmann-type equation describing evolution of the
gluon gas towards thermal equilibrium~\cite{Muell2}.

\begin{figure}[hbt]
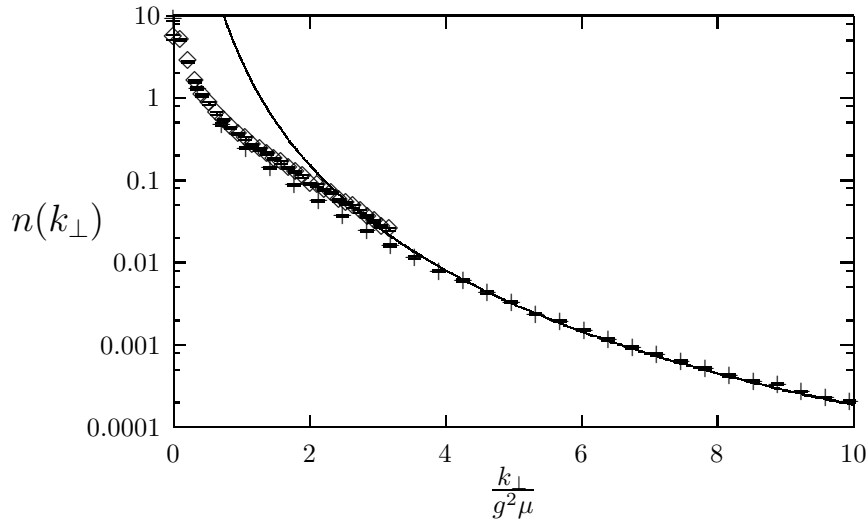

\setlength{\unitlength}{0.240900pt}
\ifx\plotpoint\undefined\newsavebox{\plotpoint}\fi
\sbox{\plotpoint}{\rule[-0.200pt]{0.400pt}{0.400pt}}%

\caption{Transverse-plane gluon number (per unit rapidity) distribution for
$g^2\mu L=35.5$ (plusses) and for $g^2\mu L=297$ (diamonds). The transverse 
momentum is expressed in units of $g^2\mu$. The solid line is a fit of the 
$g^2\mu L=35.5$ data to the perturbative expression (\ref{dpt}).}
\label{numdist}\end{figure}
The particle number is a well-defined notion in a free field theory whose 
Hamiltonian in momentum space has the form 
\begin{equation}
H_f={1\over 2}\sum_k\left(|\pi(k)|^2+\omega^2(k)|\phi(k)|^2\right),\label{hfree}
\end{equation}
where $\phi(k)$ is $k$-th momentum component of the field, $\pi(k)$ is its
conjugate momentum, and $\omega(k)$ is the corresponding eigenfrequency.
The average particle number of the $k$-th mode is then
\begin{equation}
n(k)=\omega(k)\langle|\phi(k)|^2\rangle
=\sqrt{\langle|\phi(k)|^2|\pi(k)|^2\rangle},\label{nfree}\end{equation}
where, in our case, the average $\langle\rangle$ is over the initial conditions.
Obviously, any extension of this notion to interacting theories should reduce
to the standard free-field definition of the particle number in the 
weak-coupling limit. However, this requirement alone does not define the 
particle number uniquely outside a free theory. We therefore use two different
generalizations of the particle number to an interacting theory, each having
the correct free-field limit. We verify that the two definitions agree in
the weak-coupling regime corresponding to late proper times in the central
region. We note, however, that
the theory in question may have low-lying metastable states. 
If, for a small value of $g^2\mu L$, the system finds itself in the vicinity
of a metastability, then the system is far from linearity, and both our
definitions of the number make little sense. We therefore restrict our
attention to values of $g^2\mu L$ for which energies of metastable minima
are much lower than the average energy of a configuration. In all such cases
the two definitions give results close to each other.

Our first definition is straightforward. We impose the Coulomb gauge condition
in the transverse plane: ${\vec\nabla}_\perp\cdot{\vec A}_\perp=0$ and
substitute the momentum components of the resulting field configuration into
(\ref{nfree}). At this point, there are two possibilities open to us. We can
assume $\omega(k)$ to be the standard massless (lattice) dispersion relation 
and use the middle expression of (\ref{nfree}) to compute $n(k)$.
Alternatively, we can determine $n(k)$ from the rightmost expression of
(\ref{nfree}); the middle expression of (\ref{nfree}) can then be used to obtain
$\omega(k)$.

Our second definition is based on the behavior of a free-field theory under
relaxation. Consider a simple relaxation equation for a field in real space,
\begin{equation}\partial_t\phi(x)=-\partial H/\partial\phi(x),\label{relax}
\end{equation}
where $t$ is the relaxation
time (not to be confused with real or proper time) and $H$ is the
Hamiltonian. For a free field ($H=H_f$) the relaxation equation has exactly
the same form in the momentum space with the solution
$\phi(k,t)=\phi(k,0)\,{\rm exp}(-\omega^2(k)t)$. The potential energy of the
relaxed free field is $V(t)=(1/2)\sum_k\omega^2(k)|\phi(k,t)|^2$.
It is then easy to derive the following integral expression
for the total particle number of a free-field system:
\begin{equation}
N=\sqrt{8\over\pi}\int_0^\infty{{{\rm d}t}\over\sqrt{t}}\,V(t).\label{ncool}
\end{equation}
Now (\ref{relax}) can be solved numerically for interacting fields. 
Subsequently, $V(t)$ can be determined, and $N$ can be computed by numerical
integration. Note that in a gauge theory the relaxation equations are 
gauge-covariant, and the relaxed potential $V(t)$ is gauge-invariant,
entailing gauge invariance of this definition of the particle number.
This is an attractive feature of the relaxation method. On the other hand, this
technique presently only permits determination of the total particle number
and cannot be used to find the number distribution. 

Our findings are summarized in Figures \ref{numtot} and \ref{numdist}. We 
consider these results preliminary, since we are yet to
perform a careful extrapolation to the continuum limit, as we did in the case
of the energy. In the case of the energy measurement, the systematic error 
related to a finite lattice cutoff was of the order of 10\%. For the
particle number, which is better behaved in the ultraviolet
than the energy, this systematic error should be smaller.

As Figure \ref{numtot} shows, our two definitions of the particle number agree
on a $20$\% level in a wide range of values of $g^2\mu L$, which includes the
RHIC and the LHC regimes. If we write the particle number per unit rapidity as
$N=(g^2\mu L)^2 f_N(g^2\mu L)$, then $f_N(g^2\mu L)=0.14\pm 0.03$ in that 
range.

We now estimate the number of gluons produced in one unit of rapidity, at 
central rapidities, at RHIC and LHC. We extrapolate to SU(3) in a
manner analogous to that of the energy estimate. For the physical number, 
one divides by $g^2$ and multiplies by the ratio of the relevant 
color factors ($16/9$ in this case). For $g^2\mu L \approx 116$ 
(RHIC) and $f_N=0.13$--$0.15$, we obtain $N = 778$--$897$. For 
$g^2\mu L\approx232$, (LHC) we obtain $N=3100$--$3600$. For the same 
range of $f$'s, a $g^2\mu L=150$ value for RHIC would give $N=1300$-$1500$, 
and $g^2\mu L=300$ for LHC would give $N = 5200$--$6000$. Since $N$ 
depends quadratically on $g^2\mu L$, and the latter is not known with 
with great precision, the range of the prediction is significant. What 
would be more interesting though is the slope $f$ of the ratio of the 
two, for which we have a prediction up to $20$\% at present. Varying the 
energies, and sizes of the nuclei, should enable one to extract this 
quantity. This point, and comparisons to predictions from other models, 
will be discussed further in a forthcoming paper~\cite{RVAK}.

Finally, Figure \ref{numdist} shows how $N$ is distributed among various 
momentum modes, for two extreme cases: $g^2\mu L\approx 300$ and 
$g^2\mu L\approx 35.5$. For comparison, we also show a fit of the high-momentum
tail ($k_\perp\gg g^2\mu$ of the $g^2\mu L\approx 35.5$ distribution to the 
perturbative expression (\ref{nfree}). On the low-$k_\perp$ end of the
spectrum our fully non-perturbative result deviates significantly from the 
perturbative prediction and remains finite at $k_\perp=0$. Note that the 
deviation first occurs for $k_\perp$ of the order of the non-perturbative
scale $g^2\mu$.

In summary, our numerical implementation of MV model allows one to take
into account non--perturbative effects at high parton density in the
central region. We have derived non--perturbative formulae which relate 
the energy and number of produced gluons to the gluon density and the 
size of the incoming nuclei. Varying the energy and the size of nuclei 
should enable us to test the predictive power of these formulae. 
Our treatment can be made more accurate by switching
from the SU(2) to the true physical SU(3) gauge group, by relaxing the
assumption of exact boost invariance, and by replacing periodic
boundary conditions by more realistic ones. We plan to address these
issues in future work.

\section*{Acknowledgements}
R.V.'s research was supported by DOE Contract No. DE-AC02-98CH10886. 
The authors acknowledge support from the Portuguese FCT,
under grants CERN/P/FIS/1203/98 and
CERN/P/FIS/15196/1999.

\end{document}